\documentclass[11pt]{article}
\usepackage{amssymb}
\usepackage{epsfig}

\newcommand{\BE}{\begin{equation}}
\newcommand{\EE}{\end{equation}}
\newcommand{\BA}{\begin{eqnarray}}
\newcommand{\EA}{\end{eqnarray}}

\begin{document}
\vskip 1 pt

\begin{titlepage}

\begin{center}

   {\LARGE{\bf An alternative heavy Higgs mass limit}}

\vspace*{14mm} {\Large P. Castorina$^{(1,2)}$, M. Consoli$^{(2)}$ and D. Zappal\`a$^{(2)}$}
\vspace*{4mm}\\
{$^{(1)}$Dipartimento di Fisica e Astronomia dell' Universit\`a di Catania \\
$^{(2)}$Istituto Nazionale di Fisica Nucleare, Sezione di Catania \\
Via Santa Sofia 64, I95123 Catania, Italy }
\end{center}
\begin{center}
{\bf Abstract}
\end{center}

After commenting on the present value of the Higgs particle mass
from radiative corrections, we explore the phenomenological
implications of an alternative, non-perturbative renormalization of
the scalar sector where the mass of the Higgs particle does not
represent a measure of observable interactions at the Higgs mass
scale. In this approach the Higgs particle could be very heavy, even
heavier than 1 TeV, and remain nevertheless a relatively narrow
resonance.

\vspace*{8mm}

\end{titlepage}

\section{Introduction}

The very good agreement of the Standard Model predictions with the
available experimental data (with the possible exception of neutrino
masses) strongly constrains possible frameworks for new physics. At
the same time, one essential ingredient of the theory, namely the
Higgs boson, is still missing and, at the present, there is an open
discussion to understand the marginal consistency between the result
from its direct search, $m_h
>$ 114.4 GeV (at the 95$\%$ C.L.) \cite{lep1}
and the phenomenological indication from radiative corrections
$m_h=87^{+36}_{-27}$ GeV (with the corresponding $95\%$ C.L. upper
bound $m_h < 160$ GeV \cite{lep2}), see for instance the
Introduction of Ref.\cite{barbieri}.

In general, there might be many different scenarios for $m_h$. At
the extremes of the mass range one can consider two basically
different options. The (minimal) supersymmetry, where the mass of
the lightest Higgs scalar is $m_h={\cal O}( M_w)$, and the
alternative point of view where the Standard Model, with a large
$m_h={\cal O}(1)$ TeV, represents the natural extension beyond the
massive vector boson theory. This latter point of view, where a
heavy Higgs particle would mark a second threshold for weak
interactions, was emphasized by Veltman \cite{veltman}. In his view,
a large $m_h$ represents a natural cutoff of the non-renormalizable
Glashow model \cite{glashow} just like the W mass is the natural
cutoff for the non-renormalizable 4-fermion V-A theory.

Such a heavy Higgs particle seems in contradiction with the
precision tests and thus should be ruled out. In our opinion,
however, this conclusion might be too restrictive for the following
reasons:

~~~~i) the phenomenological analysis is based on experimental
quantities where the Higgs particle mass enters only through the
radiative correction to Veltman's $\rho-$ parameter \cite{rho}.
Although for $m_h={\cal O}(1)$ TeV the known quadratic two-loop
contribution remains considerably smaller than the logarithmic
one-loop term, the authors of Ref.\cite{rho2}, from the expected
magnitude of the higher-order terms, were estimating that
perturbation theory might break down when $m_h$ is in this region
("It is very unclear what happens if the Higgs mass is in this
range"). For this reason, strictly speaking, finite-order
calculations in the Higgs sector might become inconsistent for
$m_h={\cal O}(1)$ TeV and large values of $m_h$ be still
phenomenologically viable.

~~~~ii) one cannot exclude that, if the physical spectrum is richer
than that expected in the Standard Model, see e.g.
Refs.\cite{peskin,barbieri}, even for a large $m_h$, the resulting
effective $S$ and $T$ parameters might mimic those perturbatively
computed for small values of $m_h$.

~~~~iii) the quality of the global fit, when including {\it all}
electroweak data, $\chi^2/d.o.f.$=29.7/15 with a C.L. 1.3$\%$
\cite{chi}, indicates that some subsets of data are marginally
consistent with each other. We emphasize that some of them point to
a large $m_h$. In this sense, the present reading of the data is not
completely free of ambiguities.

~~~~iv) the measured hadronic cross sections at LEP2 show an average
excess of about $+1.5\%$ with respect to the SM prediction
\cite{alphas}. Due to the known positive $(m_h,\alpha_s)$
correlation in the radiative corrections, see e.g. \cite{hioki},
extrapolating this effect back on the Z peak would tend to increase
considerably the fitted value of $m_h$ from the electroweak data.

On the basis of the previous arguments, it might be worth to
re-consider from scratch the case of a heavy Higgs particle ${\cal
O}(1)$ TeV and the possible phenomenology expected at LHC. Radiative
corrections represent then a separate issue. In the end, if the
Higgs particle will turn out to be heavy and if consistency with
radiative corrections will still require the introduction of new
physics, experimental evidences for the needed new physical sector
might be found within the same LHC energy range.

Here the basic point we want to emphasize is that a large $m_h$ does
not necessarily imply the presence of strong interactions at the
Higgs mass scale. In fact, one can consider an alternative,
non-perturbative renormalization of the scalar sector where this
conclusion is not true. In this approach, the Higgs particle could
be very heavy, even heavier than 1 TeV, and remain nevertheless a
relatively narrow resonance.

In the following, we shall first review in Sect.2 the standard
picture associated with a heavy Higgs particle. Then, in Sect.3, we
shall illustrate an alternative description of spontaneous symmetry
breaking in $\lambda\Phi^4$ theories where the mass parameter $m_h$
does not represent by itself a measure of any observable
interaction. Further, in Sect.4, we shall explore the possible
implications of this picture for the Standard Model. Finally, in
Sect.5, we shall present a summary and our conclusions.

\section{The standard heavy Higgs mass limit}

To start with, let us preliminarily observe that the existence of a
non-perturbative regime for large $m_h$ is usually deduced in two
very different ways. On the one hand, without considering the
self-interactions in the Higgs sector and by simply using
leading-order calculations in the unitary gauge, one finds
\cite{lee} the violation of tree-level unitarity in the scattering
amplitudes for longitudinal polarized vector bosons for values of
$m_h$ such that \BE \label{unit} {{3m^2_h}\over{16\pi v^2_R}}\sim
1\EE $v_R$ being the physical vacuum expectation value of the scalar
field $v_R \sim {{1}\over{\sqrt{G_F \sqrt{2}}}}\sim 246$ GeV. On the
other hand, by explicitly considering the Higgs sector, the previous
strong interactions emerge as a remnant of the scalar dynamics at
zero gauge coupling. In fact, the Higgs field is predicted to be
self-interacting with a tree-level, contact strength
\BE\label{lambda}\lambda^{(1)}= {{3m^2_h}\over{v^2_R}}\EE so that
Eq.(\ref{unit}), equivalent to $m_h\sim 1$ TeV, can also be
expressed as $ {{\lambda^{(1)}}\over{16\pi}}\sim 1$. The link
between the two descriptions, in the high-energy limit $M_w \ll
\sqrt{s}$, represents the content of the so called "Equivalence
Theorem" \cite{equivalence}. In this context, the value $m_h \sim 1$
TeV does not represent an upper limit but, rather, indicates a mass
region where one should expect substantial non-perturbative effects.

Beyond the tree-approximation, in the usual approach, the
self-coupling $\lambda^{(1)}$ in Eq.(\ref{lambda}) no longer
represents the contact interaction but is now interpreted as a sort
of boundary condition for the running coupling at a scale $\mu\sim
m_h$. To explore the implications of this identification, let us
denote in general by $\lambda_s$ the contact interaction at some
large energy scale $\Lambda_{\rm s}$  that, in principle, in a
quantum field theory one might also decide to send to infinity. Let
us also denote by $\lambda^{(2)}$ the running coupling at the scale
$\mu=m_h \ll \Lambda_s$.

By assuming the perturbative $\beta-$function of $\lambda\Phi^4$
theory \BE \label{betapert} \beta_{\rm pert}(\lambda)=
{{3\lambda^2}\over{16\pi^2}} +{\cal O}(\lambda^3) \EE one gets the
relation \BE \label{scale} \ln {{m_h}\over{\Lambda_s}}~=~ \int
^{\lambda^{(2)} }_{\lambda_s}~{{dx}\over{\beta_{\rm pert}(x)}}\EE
Therefore, when $\Lambda_s/m_h \to \infty$ one ends up with an
infinitesimal low-energy coupling \BE \label{leading}
\lambda^{(2)}\sim {{{\rm 16 \pi^2}}\over{3\ln
{{\Lambda_s}\over{m_h}} }}~ =~ \epsilon\EE regardless of how large
$\lambda_s$ might be.

The identification $\lambda^{(1)}\sim \lambda^{(2)}$ leads to values
of $m_h$ that have to decrease by increasing the magnitude of
$\Lambda_s$. In particular, if the maximal Higgs mass and the lowest
degree of locality of the theory are connected through the order of
magnitude relation $\ln {{(\Lambda_s)_{\rm min}}\over{(m_h)_{\rm
max}}}\sim 1$, one obtains \BE {{ 16\pi^2}\over{3}}\sim
{{3(m^2_h)_{\rm max}}\over{v^2_R}} \EE and the order of magnitude
estimate $(m_h)_{\rm max} \sim 1$ TeV \cite{lattice}. At the same
time, if very heavy, say 800-1000 GeV, one should detect some
strongly interacting sector at LHC by measuring some peak cross
section that is proportional to $m_h$. In this mass range, the Higgs
particle is expected to be a very broad resonance with a decay width
into longitudinal W's\BE\label{width} \Gamma(h \to W_L W_L )\sim {{3
m^3_h}\over{32\pi v^2_R}} \EE that is comparable to its mass.

On the other hand, by still accepting a vanishingly small low-energy
interaction, as in Eq.(\ref{leading}) when $\Lambda_s \to \infty$,
there might be alternative frameworks where the standard assumption
${\lambda^{(1)}}\sim {\lambda^{(2)}}$ is {\it not} true. This
happens in the approach of Refs.\cite{cs1,cs2} where the two
couplings in Eq.(\ref{lambda}) and Eq.(\ref{leading}) have a
qualitatively different meaning. Namely, the coupling in
Eq.(\ref{lambda}) emerges as a {\it collective} self-interaction of
the scalar condensate that enters the effective potential. The
coupling in Eq.(\ref{leading}), on the other hand, is appropriate to
describe the {\it individual} interactions among the elementary
excitations of the vacuum, i.e. the singlet Higgs field $h(x)$ and
the Goldstone boson fields $\chi_i(x)$. Thus, the continuum limit
$\Lambda_s \to \infty$ would have a finite $m_h/v_R$ ratio but
${\lambda^{(2)}} \to 0$ with trivially free Higgs and Goldstone
boson fields.

\section{An alternative picture of symmetry breaking}

Before addressing the Standard Model, we shall first remind the
basic results of Refs.\cite{cs1,cs2}. These were obtained by
exploring the structure of the effective potential in that
particular class of approximations (gaussian and postgaussian
approximations both for the discrete-symmetry and
continuous-symmetry O(N) theory \cite{rit2}) where the scalar
self-interaction effects can be reabsorbed into the parameters of an
effective quadratic hamiltonian.

The key observation is that, within the generally accepted
"triviality" of the theory in 3+1 dimensions, and by retaining an
infinitesimal 2-body coupling of the type in Eq.(\ref{leading}), one
can nevertheless obtain a non-trivial effective potential $V_{\rm
eff}$. In fact, when $\Lambda_s \to \infty$, ``triviality'' dictates
the vanishing of all observable, i.e. non-zero momentum, scattering
processes. Thus it does not forbid a non-trivial zero-momentum
dynamics, that enters the effective potential, provided that its
effects can be re-absorbed into the first two moments of a gaussian
structure of Green's functions. After reviewing the original
argument, we shall present the very same physical conclusions from a
different point of view that, perhaps, may sound more familiar to
most readers.

To illustrate the basic point, let us consider the familiar one-loop
potential, as originally computed by Coleman and Weinberg \cite{CW}
\BE \label{oneloop} V^{\rm 1-loop}_{\rm eff}(\varphi)= {{\lambda
\varphi^4}\over{4!}} + {{\lambda^2\varphi^4}\over{256\pi^2}}\left(
\ln {{\lambda\varphi^2}\over{2\Lambda^2_s}}-{{1}\over{2}}\right) \EE
Such basic calculation predicts a weakly first order phase
transition. In fact, at one loop, the version of the theory
corresponding to the mass renormalization condition $V''_{\rm
eff}(\varphi=0)=0$ gives non-trivial absolute minima for values
$\varphi=\pm v$ such that \BE \label{mass} m^2_h={{\lambda
v^2}\over{2}}=\Lambda^2_s\exp -{{32\pi^2}\over{3\lambda}} \EE This
yields the equivalent form of the effective potential \BE V^{\rm
1-loop}_{\rm eff}(\varphi)=
{{\lambda^2\varphi^4}\over{256\pi^2}}\left( \ln
{{\varphi^2}\over{v^2}}-{{1}\over{2}}\right) \EE and of the vacuum
energy
 \BE {\cal E}=V^{\rm 1-loop}_{\rm eff}(v)=-
{{m^4_h}\over{128\pi^2}} \EE Notice that Eq.(\ref{mass}) gives the
relation $\lambda\sim \epsilon $ for the coupling constant as in
Eq.(\ref{leading}) so that the quadratic shape at the minima
${\varphi=\pm v}$ \BE {{d^2 V^{\rm 1-loop}_{\rm
eff}}\over{d\varphi^2}}= {{\lambda^2 v^2}\over{32\pi^2}} \sim
\epsilon m^2_h\EE is infinitesimal in units of $m_h$.

As discussed by Coleman and Weinberg, one may object to such a
straightforward minimization procedure. In fact, the
``renormalization-group-improved'' result \BE V^{\rm LL}_{\rm
eff}(\varphi)\sim
{{\lambda(\mu^2=\lambda\varphi^2)}\over{4!}}~\varphi^4 \EE obtained
by resumming leading-logarithmic terms to all orders, gives no
non-trivial minima and confirms the expectations based on a
second-order phase transition, as at the classical level. The
conventional view is that the latter improved calculation is
trustworthy while the former is not, the argument being that the
minimum occurs where the one-loop correction is as large as the
classical tree-level term. However, there is an equally strong
reason to distrust the RG-improved result \cite{cs2}: it amounts to
re-summing a geometric series of leading logarithmic terms that is
actually a divergent series. The moral is that one cannot trust
perturbation theory, improved or not, and one needs an alternative
approximation scheme.

The gaussian approximation to the effective potential, being of
variational nature, gives one possible clue \cite{gaussian,cs1}. At
the same time, by accepting the ``triviality'' of  $\Phi^4$ theories
in 3+1 space-time dimensions, it should be reliable when taking the
continuum limit $\Lambda_s \to \infty$ where all higher order
Green's functions should be expressible in terms of the first two
moments of a gaussian distribution. Now, the gaussian effective
potential is in complete agreement with the basic one-loop result
(\ref{oneloop}) thus confirming the indications of a weakly
first-order phase transition. Moreover, the same conclusion persists
in the more elaborate postgaussian approximations both for the
discrete-symmetry and continuous-symmetry O(N) theory \cite{rit2}.

The point is that, differently from the leading-log resummation, the
alternative infinite set of gaussian and post-gaussian corrections
simply redefines the coupling entering the tree-level potential and
the mass entering the zero-point energy, thus preserving the basic
one-loop structure in Eq.(\ref{oneloop}). In the gaussian
approximation the relevant relations can be given in the reasonably
compact form as \BE \label{gaussian} V^{\rm G}_{\rm eff}(\varphi)=
{{\hat{\lambda} \varphi^4}\over{4!}} +
{{\Omega^4(\varphi)}\over{64\pi^2}}\left( \ln
{{\Omega^2(\varphi)}\over{\Lambda^2_s}}-{{1}\over{2}}\right) \EE
with \BE \hat{\lambda}= {{\lambda}\over{1+
{{\lambda}\over{16\pi^2}}\ln {{\Lambda_s}\over{ \Omega(\varphi) }}
}} \EE the variational mass $\Omega(\varphi)$ being determined
self-consistently as $\Omega^2(\varphi)= \hat{\lambda}\varphi^2/2$
with the minimum-value relation $m_h=\Omega(v)$.

In this sense, the one-loop potential admits a non-perturbative
interpretation being the prototype of all ``triviality-compatible''
approximations to $V_{\rm eff}$. The basic results that are found in
this class of approximations are therefore the same as at one loop,
namely

i) the physical mass of the shifted Higgs field is given by \BE
\label{intu} m^2_h\sim \epsilon v^2\EE ~~~~ii) the vacuum energy
density scales as\BE {\cal E} \sim - m^4_h \EE In the previous
relations $\varphi$ indicates the bare vacuum expectation value of
the scalar field, taken as a variational parameter, and $v$ its
stationarity value. Notice that, by introducing the critical
temperature $T_c$ at which symmetry can be restored (by the  order
of magnitude relation $|{\cal E}|\sim T^4_c$), one finds that $T_c$
is finite in units of $m_h$.

Now, adopting $m_h$ as the unit mass scale, while the vacuum energy
density itself, $|{\cal E}|\sim m^4_h$, is finite, the {\it slope}
of the effective potential has necessarily to be infinitesimal \BE
V''_{\rm eff}(\varphi=v)\sim {{m^4_h}\over{v^2}}\sim \epsilon
m^2_h\EE because, starting from $\varphi^2=0$, the finite vacuum
energy ${\cal E}$ will be attained for the infinitely large value
$\varphi^2= v^2\sim m^2_h/\epsilon$. Thus, if one wants to introduce
a definition of the vacuum field, say $\varphi_R=v_R$, to match the
quadratic shape of the effective potential to the physical mass,
i.e. \BE \label{match} V''_{\rm eff}(\varphi_R=v_R)=m^2_h \EE one
needs a potentially divergent re-scaling for the bare vacuum field
\BE \varphi^2=Z_\phi \varphi^2_R\EE with \BE Z_\phi={\cal
O}(1/\epsilon)\EE since \BE \label{match2} V''_{\rm
eff}(\varphi_R=v_R)=Z_\phi V''_{\rm eff}(\varphi=v)\EE The matching
condition Eq.(\ref{match}), that merely expresses the consistency
requirement that the renormalized zero-momentum two-point function
$V''_{\rm eff}(v_R)$ gives the $p_\mu \to 0$ limit of the inverse
connected propagator $(p^2+m^2_h)$, is implicit when expanding in
the broken phase the full scalar field $\Phi(x)$ into the sum of a
vacuum part $v_R$ and of a free-field-like quantum fluctuation
$h(x)$ with a given physical mass $m_h$. The generalization to the
continuous-symmetry case, which is relevant for the Standard Model,
is straightforward and consists in expressing the isodoublet scalar
field as \BE \label{doublet} \Phi(x)= e^{i \theta_a (x)
\sigma_a}~{{1}\over{\sqrt{2}}}~(0, ~v_R + h(x) ) \EE by defining the
Goldstone boson fields $\chi_a(x)$ through the relation
$\theta_a(x)= \chi_a(x)/v_R$.

For completeness, we want to mention that the standard perturbative
calculations of $V_{\rm eff}$, while still maintaining  $m^2_h\sim
\epsilon v^2$, provide however the different scaling law for the
vacuum energy density ${\cal E}\sim -\epsilon v^4$. Thus, in this
other calculation scheme, one finds $V''_{\rm eff}(v)\sim m^2_h$ and
$v\sim v_R$. Since the vacuum energy density ${\cal E}\sim
-{{m^4_h}\over{\epsilon}}$ diverges in units of $m^4_h$ when
$\epsilon \to 0$, the critical temperature $T_c\sim
{{m_h}\over{\epsilon^{1/4}}}$ now diverges in units of $m_h$.

Returning to the picture of Refs.\cite{cs1,cs2}, we observe that the
introduction of the potentially divergent re-scaling $Z_\phi$ does
not violate any rigorous result since the scalar condensate is not a
quantum field that enters the asymptotic representation for the
S-matrix. Due to $Z_\phi$, $m_h$ and $v_R$ scale uniformly in the
continuum limit where $\epsilon \to 0$ and therefore the coupling
$\lambda^{(1)}$ emerges as the natural collective interaction of the
scalar condensate with itself \BE\epsilon Z_\phi\sim
{{m^2_h}\over{v^2}}~ {{v^2}\over{v^2_R}} \sim \lambda^{(1)} \EE On
the other hand, due to ``triviality'', the shifted quantum fields
have necessarily to undergo a trivial re-scaling $Z_h =1 +{\cal
O}(\epsilon)$. For this reason, at low-energy, they remain weakly
interacting entities with typical strength $\lambda^{(2)} \sim
\epsilon$, as in Eq.(\ref{leading}). The difference $Z_\phi \neq
Z_h$, or $\lambda^{(1)} \neq \lambda^{(2)}$, reflects the basic
phenomenon of vacuum condensation \cite{old} in a ``trivial" theory
and would be sensitive to the presence of a very large cutoff
$\Lambda_s$.

From this point of view, the situation is similar to the phenomenon
of superconductivity in non-relativistic solid state physics. There
the transition to the new, non-perturbative phase represents an
essential instability that occurs for any infinitesimal two-body
attraction between the two electrons forming a Cooper pair. At the
same time, however, the energy density of the superconductive phase
and all global quantities of the system (energy gap, critical
temperature,...) depend on a much larger {\it collective} coupling
obtained after re-scaling the tiny 2-body strength by the large
number of states near the Fermi surface.

Notice also that $Z_\phi$ enters the extraction of the physical
$v_R$ from the bare $v$ measured in lattice simulations. In this
way, one will not run in contradiction with the existing data where
a single wave function renormalization constant $Z=Z_h \sim 1$ has
always been assumed both for the condensate and the fluctuation
field \cite{ccc,stevenson}. We emphasize that even the authors of
the critical response paper of Ref.\cite{balog} must admit that
``the unconventional picture of symmetry breaking cannot be ruled
out by present numerical simulations".

For a complete description of the $v_R-m_h$ interdependence, we
summarize below the main quantitative results. These can be
conveniently expressed in terms of the parameter pair
$(\zeta,v^2_R)$ that conveniently replaces the bare mass and bare
coupling. The parameter $\zeta$, with $0 < \zeta \leq 2$, is defined
by the relation \cite{cs2} \BE \label{zeta} V''_{\rm
eff}(\varphi_R=v_R)= m^2_h=8 \pi^2 \zeta v^2_R \EE with the upper
bound $\zeta =2$ coming from vacuum stability since one finds
\BE{\cal E}=-{{\pi^2}\over{2}}\zeta(2-\zeta)v^4_R \EE Also, in terms
of $\zeta$ the condensate re-scaling factor is \BE \label{zetazeta}
  Z_\phi= 3\zeta\ln {{\Lambda_s}\over{m_h}} \EE
Finally, one obtains \BE \label{zeta4} V''''_{\rm
eff}(\varphi_R=v_R)= {{3m^2_h}\over{v^2_R}}~(1+ {{m^2_h}\over{3
\pi^2 v^2_R}})\EE As shown in Ref.\cite{stancu}, in the heavy-Higgs
mass region $\zeta \sim 1$, corresponding to $m_h={\cal O}(1)$ TeV,
the inclusion of vector bosons introduces only small corrections
${\cal O} (M^2_w/m^2_h)$ with respect to the relations
(\ref{zeta})$-$(\ref{zeta4}) of the pure scalar case.

As anticipated, we conclude this part by mentioning a more
conventional way to understand the crucial relation $\lambda^{(1)}
\neq \lambda^{(2)}$. With an alternative, non-perturbative stability
analysis of the theory, we have found that the quantity
$\lambda^{(1)}\sim m^2_h/v^2_R$ is a collective self-coupling of the
condensate and does not describe the low-energy interactions of the
Higgs field and of the Goldstone bosons for which the relevant
coupling is rather $\lambda^{(2)}={\cal O}(\epsilon)$. It is
possible to characterize this alternative approach without
mentioning the effective potential and by only considering the
possible choices of boundary conditions in Eq.(\ref{scale}). In
fact, one could say that standard perturbation theory, assuming
$\lambda^{(1)} \sim \lambda^{(2)}$, is limited to parameter pairs
$(m_h,\Lambda_s)$ for which $m^2_h/v^2_R$ and $\ln(\Lambda_s/m_h)$
{\it cannot be}, at the same time, very large. Now, suppose that one
would try to relax this condition and describe the complementary
situation where $m^2_h/v^2_R$ and $\ln(\Lambda_s/m_h)$ {\it can be},
at the same time, very large. In this case, when starting from
values $ {{\lambda^{(1)}}\over{16\pi}}\sim 1$, it becomes natural to
associate the range of values $\lambda(\mu)\sim \lambda^{(1)}\gg 1$
with an {\it ultraviolet} scale $\mu \gg m_h$ rather than with
region $\mu \sim m_h$ where $\lambda(\mu)\sim \lambda^{(2)}={\cal
O}(\epsilon)$. In this alternative reading, abandoning
$\lambda^{(1)}$ and using $\lambda^{(2)}$, to describe the
low-energy interactions of the Higgs field and of the Goldstone
bosons, is nothing but the standard renormalization-group evolution
from $\lambda^{(1)}$ down to $\lambda^{(2)}$.

Thus, our picture is relevant to that corner, where both
$m^2_h/v^2_R$ and $\ln(\Lambda_s/m_h)$ might be much larger than
unity, that does not exist in the usual approach \cite{corner}. In
the next section, we shall explore the phenomenology expected in
this other version of the theory where $m_h={\cal O}(1)$ TeV and
$\Lambda_s$ can be as large as the highest energy boundaries that
are usually considered, namely the grand-unification scale
$\Lambda_{\rm gauge}\sim 10^{15}$ GeV or even the Planck scale.

\section{A new heavy Higgs phenomenology}

To explore the phenomenological implications of the alternative
picture of symmetry breaking described in Sect.3, we shall first
analyze the physical meaning of the Equivalence Theorem, as
explained in Ref.\cite{bagger}. According to these authors, in any
renormalizable $R_\xi$ gauge, the limit $g=g_{\rm gauge} \to 0$ is
{\it smooth}. This leads to predict that, in the full Standard Model
where $g\neq 0$, there can be no physical interaction (say a peak
cross section, a decay width,..) that grows proportionally to $m_h$
{\it unless} the same coupling is already present in the theory at
$g= 0$. With these premises, the scattering amplitudes for
longitudinal vector boson scattering in the $g\neq 0$ theory can be
obtained from the corresponding amplitudes for the scattering of the
Goldstone boson fields $\chi_i$ in the $g= 0$ theory. Namely, in the
high-energy limit $M_w \ll \sqrt{s}$, one can establish the
following relation between the two T-matrix elements (the S-matrix
is $S=1+iT$) \BE T(W_L W_L \to W_LW_L)=C^4 T_{g=0}(\chi\chi \to \chi
\chi) \EE where \cite{bagger} \BE C=1+{\cal O}(g^2)\EE and, from now
on, we omit for simplicity the index $i$ both in the vector boson
and Goldstone boson fields.

Since the Equivalence Theorem is valid to lowest order in $g^2$ but
to {\it all} orders in the scalar self-coupling one expects it to
remain valid even if the scalar sector were treated non
perturbatively.

To illustrate explicitly  the basic point, let us consider for
simplicity, as in Ref.\cite{yndurain}, the case of a pure SU(2)
symmetry, i.e. setting $\sin\theta_w=0$, with an invariant
Lagrangian given by  $$ {\cal L}_{\rm inv}=
-{{1}\over{4}}G^a_{\mu\nu}G^a_{\mu\nu}-{{1}\over{2}}M^2_wW^2-
{{1}\over{2}}(\partial_\mu h)^2-
 {{1}\over{2}}\partial_\mu\chi^aD_\mu\chi^a+
{{1}\over{2}}gW^a_\mu(h\partial_\mu\chi^a-\chi^a \partial_\mu h)$$
$$-M_w\chi^a\partial_\mu W^a_\mu-{{1}\over{2}}gM_wW^2h
-{{1}\over{8}}g^2W^2(h^2+\chi^a\chi^a)-U_{\rm Higgs}+... $$ and a
fourth-order Higgs potential \BE \label{uhiggs} U_{\rm Higgs}=
{{1}\over{2}}m^2_h h^2 +\epsilon_1 rgM_w ~h(\chi^a\chi^a +h^2)
+{{1}\over{8}}\epsilon_2rg^2~(\chi^a\chi^a +h^2)^2\EE with \BE
r={{m^2_h}\over{4M^2_w}}\EE This structure is valid to all orders in
the scalar self-interactions but to lowest order in $g^2$. The dots
indicate higher-order potential and derivative terms in the
low-energy expansion of the full effective action for the scalar
fields, evaluated in gaussian and post-gaussian approximations and
renormalized as in Sect.3 by imposing the condition $V''_{\rm
eff}(v_R)=m^2_h$ . The two phenomenological parameters $\epsilon_1$
and $\epsilon_2$ should be set to unity in ordinary perturbation
theory. In general, substantial differences from unity would signal
new physics in the scalar sector or, as in our case, indicate an
alternative description of symmetry breaking where
$\epsilon^2_1=\epsilon_2=1/Z_\phi\sim \epsilon$.

Concerning the high-energy scattering of longitudinal W's, we
observe that, at the tree-level, it can be computed directly as in
Ref.\cite{lee}. In this case, one starts from a tree-level amplitude
which is formally ${\cal O}(g^2)$. However, after contracting with
the longitudinal polarization vectors $\sim {{k_\mu}\over{M_w}}$ on
the external legs, one ends up, for energies $\sqrt{s} \to \infty$,
with the same contact coupling $\lambda^{(1)}$ as for the scalar
case. Thus longitudinal W's interact precisely with a tree-level,
contact strength \BE \lambda^{(1)}\sim 3g^2 {{m^2_h}\over{4 M^2_w}}
\EE Here the factor $g^2$ comes from the vertices. The factor
$m^2_h$ emerges, in the infinite-energy limit, after combining with
the other contributions the graphs with the Higgs propagator.
Finally the $1/M^2_w$ derives from contracting with the longitudinal
polarization vectors. The identification with $\lambda^{(1)}$ only
assumes the Standard Model relation $M^2_w={{g^2v^2_R}\over{4}}$
(valid up to few percent corrections such as those associated with
the renormalization of the fine structure constant from zero energy
to the W-mass scale). By taking into account $m^2_h/s$ terms (but
neglecting $M^2_w/s$ and $M^2_w/m^2_h$) the tree-level amplitudes
can be conveniently expressed as \cite{lee} \BE \label{tree}
a_0(\chi \chi \to \chi \chi)\sim a_0(W_L W_L \to W_L W_L)=-
{{\lambda^{(1)} }\over{48 \pi}}~ f (s/m^2_h) \EE with \BE f(x)= 2~
-{{1}\over{1-x}}~ -~{{1}\over{x}} ~\ln (1+x) \EE Let us now consider
the effect of loop corrections in the low-energy region $\sqrt{s}
\lesssim m_h$. In this case, with a Higgs potential as in
Eq.(\ref{uhiggs}), one finds a T-matrix element at $g=0$ \BE
T_{g=0}(\chi\chi \to \chi \chi)=O(\epsilon)\EE As discussed at the
end of Sect.3, this result is equivalent to use the $\beta-$function
in Eq.(\ref{betapert}) to re-sum higher order effects in $\chi \chi$
scattering. Thus one is driven to deduce a similar result for
longitudinal W-scattering, namely  \BE T(W_L W_L \to W_LW_L)=(1
+{\cal O}(g^2)) ~T_{g=0}(\chi\chi \to \chi \chi)=O(\epsilon) \EE The
power of the Equivalence Theorem is that, knowing the $\chi\chi$
amplitude, one does not need to re-sum the analogous troublesome
longitudinal W-scattering graphs to all orders. For $\Lambda_s \to
\infty$, and whatever the tree-level high-energy coupling, they will
always interact with a coupling ${\cal O}( \epsilon)$ at a scale
$\sqrt{s} \lesssim m_h$. To the leading-logarithmic accuracy, this
means to replace in Eq.(\ref{tree}) the tree-level coupling
$\lambda^{(1)}$ with $\lambda^{(2)}$ thus obtaining \BE \label{anew}
a(W_L W_L \to W_L W_L)\sim a(\chi \chi \to \chi \chi)\sim -
{{\pi}\over{9 \ln {{\Lambda_s}\over{m_h}} }}~ f (s/m^2_h) \EE Just
to have an idea, for $m_h\sim$ 2 TeV the coefficient of  $f
(s/m^2_h)$ in Eq.(\ref{tree}) is about 1.3. On the other hand, for
$\Lambda_s \sim 10^{15}$ GeV, the coefficient in Eq.(\ref{anew}) is
about 0.013. In this way, both the intermediate energy range $M_w
\ll \sqrt{s} \ll m_h$, where $f \sim -s/m^2_h$, and the resonance
region $s\sim m^2_h$ correspond to a weak-coupling regime. Finally,
in the region $ m_h\ll \sqrt{s} \ll \Lambda_s $, where $f\sim 2$,
one should replace $m_h$ with $\sqrt{s}$ within the logarithm in
Eq.(\ref{anew}).

Analogously, for the Higgs decay width one can write \BE \Gamma(h
\to W_LW_L)=(1 +{\cal O}(g^2)) ~ \Gamma_{g=0}(h \to \chi \chi) \EE
By expressing the decay width into Goldstone bosons as \BE
\label{width2} \Gamma_{g=0}(h \to \chi \chi)\sim ~{{3\epsilon^2_1
m^2_h}\over{32\pi v^2_R}}~ m_h \EE one finds for $\epsilon^2_1\sim
1$ the conventional result and for $ \epsilon^2_1 \sim \epsilon $
our alternative description. Thus, in this latter approach, where
the ratio \BE \gamma={{\Gamma_{g=0}(h \to \chi \chi)}\over{m_h}}=
{\cal O}(\epsilon) \EE the Higgs particle could be very heavy and
still remain a relatively narrow resonance. As a numerical estimate,
by using eq.(\ref{zetazeta}) with $\Lambda_s \sim 10^{15}$ GeV and
$m_h\sim$ 2 TeV (so that $\zeta\sim 1$), one obtains
$\epsilon^2_1={{1}\over{Z_\phi}}\sim 10^{-2}$ and \BE \label{width4}
\gamma={{\Gamma_{g=0}(h \to \chi \chi)}\over{m_h}}\sim 0.024 \EE
Notice that both $ \lambda^{(2)}$ and $\gamma$ vanish when $\epsilon
\to 0$. Therefore, the only energy region where one can get a non
trivial $\chi\chi$ scattering is within the zero-measure set
${{\Delta \sqrt{s}}\over{ m_h}}\sim \gamma$ around the Higgs
resonance where the amplitude is $\sim
{{\lambda^{(2)}}\over{\gamma}}$.

In conclusion, by renormalizing the pure scalar sector as discussed
in Sect.3, one is driven to deduce that in the $g \neq 0$ theory the
Higgs decay width can only acquire new contributions proportional to
$g^2 m_h\ll m_h$. For a large $m_h$, a decay width proportional to
${{ m^3_h}\over{8\pi v^2_R}}$, as well as the existence of peak
cross sections proportional to ${{ m^2_h}\over{8\pi v^2_R}}$, would
imply the existence of large observable interactions at $s \sim
m^2_h$ (for $g=0$) while, according to ``triviality'', for
$\Lambda_s \to \infty$ there can be none.

We have also found that, at the leading logarithmic accuracy, by
assuming the perturbative $\beta-$function of $\lambda\Phi^4$ theory
to describe the evolution of the coupling in $\chi \chi$ scattering,
the Equivalence Theorem requires the replacement $\lambda^{(1)} \to
\lambda^{(2)}$ as the correct strength to describe longitudinal W's
scattering at scales $\sqrt{s} \lesssim m_h$.

As anticipated, this picture is just the opposite of the usual
perturbative perspective based on the identification
$\lambda^{(1)}\sim \lambda^{(2)}$. There, for a large $m_h={\cal
O}(1)$ TeV, the lowest degree of locality of the theory is set by
the Higgs sector that at a scale $\Lambda_s\sim m_h\ll \Lambda_{\rm
gauge}$ has to be replaced by new degrees of freedom. Here,
independently of $m_h$, there is no problem in taking the
$\Lambda_s\to \infty$ limit. Thus one might be driven to reverse the
hierarchical relation as if, in this alternative approach,
$\Lambda_s$ might be even larger than $\Lambda_{\rm gauge}$ (say
$\Lambda_s\sim 10^{19}$ GeV).

\section{Summary and conclusion}

Summarizing: in this paper, motivated by the stability analysis of
 the theory in a class of approximations to the effective potential
 that are consistent with the basic triviality property, we have
 explored the phenomenological implications of relaxing the standard
 equivalence between Eq.(\ref{lambda}) and Eq.(\ref{leading}).
 Namely, the coupling $\lambda^{(1)}=  {{ 3 m^2_h}\over{v^2_R}}$ in
 Eq.(\ref{lambda}) emerges as a {\it collective} self-interaction of
 the scalar condensate and can remain finite in the continuum limit.
 The coupling $\lambda^{(2)}= {\cal O}(\epsilon)$ in
 Eq.(\ref{leading}), on the other hand, is appropriate to describe
 the {\it individual} low-energy interactions of the elementary
 excitations of the vacuum, i.e. the singlet Higgs field $h(x)$ and
 the Goldstone bosons $\chi_i(x)$, and has to vanish when $\Lambda_s
 \to \infty$. Therefore a heavy Higgs might not necessarily imply
 observable strong interactions at the Higgs mass scale.

 Similar considerations apply to longitudinal W's. In fact, from the
 standard tree-level calculations, and by using the Equivalence
 Theorem, one obtains a contact coupling $\lambda_s=\lambda^{(1)}$
 for $\sqrt{s}\to \infty$. Analogously, at the level of loop
 corrections, by assuming a $\beta-$function as in
 Eq.(\ref{betapert}) to compute the downward evolution of the scalar
 self-coupling, the same Equivalence Theorem implies that at low
 energies $\sqrt{s} \lesssim m_h$ longitudinal W's will interact with
 an observable coupling $\lambda^{(2)}={\cal O}(\epsilon)$,
 regardless of how large their contact coupling $\lambda_s$ may be.

 This type of conclusion applies to all sectors of the theory where
 the ratio $m^2_h/v^2_R$ is assumed to represent a direct measure of
 observable, low-energy interactions, and thus, for instance, to the
 structure of the radiative corrections to the $\rho-$parameter
 beyond one-loop level. A general analysis of the problem requires to
 separate out preliminarily the contributions where $m_h$ enters as a
 genuine mass parameter from those where it is traded for a coupling
 constant entering the scalar 3- and 4-point functions $\Gamma_3$ and
 $\Gamma_4$ that describe the self-interactions of the Higgs and
 Goldstone boson fields. By adopting the same notations as in Sect.4,
 one can express the low-energy boundary conditions in the general
 form $\Gamma_3={{ 3\epsilon_1 m^2_h}\over{v_R}}$ and $\Gamma_4={{
 3\epsilon_2 m^2_h}\over{v^2_R}}$ where $m_h$ and $v_R$ are defined
 through the effective potential as $V''_{\rm
 eff}(\varphi_R=v_R)=m^2_h$.

 Now, in the standard perturbative approach, where there is no
 distinction between the collective self-interactions of the
 condensate and those of the fluctuating fields, one is forced to fix
 $\epsilon_1=\epsilon_2=1$, and thus there will be corrections
 proportional to $(m^2_h/v^2_R)^n$ at the $(n+1)$ loop level. In our
 alternative scheme, on the other hand, the $\epsilon_i$ parameters
 vanish when $\Lambda_s\to \infty$ and thus the leading contribution
 to $\Delta \rho$ from the Higgs sector would simply consist of the
 logarithmic one-loop term.

 As previously observed, however, our description quantitatively
 differs from the usual one only for large $m_h$ where, in the
 presence of a very large cutoff $\Lambda_s$, we predict a relatively
 narrow Higgs decay width. If $m_h$ is small, the existence or not of
 observable effects that are proportional to powers of $m^2_h/v^2_R$
 becomes irrelevant. For this reason, if in the end the present
 indication $m_h={ \cal O}(100)$ GeV will be confirmed, we cannot see
 any definite signal to distinguish the two descriptions of symmetry
 breaking. Our alternative point of view may still represent a
 logical possibility but there would be no more compelling
 theoretical reason that can be traced back to the effective nature
 of the standard perturbative approach for large $m_h$.


\begin{thebibliography} {99}
\bibitem{corner}
The distinction between $\lambda^{(1)}$ and $\lambda^{(2)}$ becomes
relatively unimportant for small values of $m^2_h/v^2_R$ when both
couplings are in the perturbative domain. In this other limit, the
phenomenological differences with the standard perturbative approach
are modest.
\end{thebibliography}
\end{document}